\documentstyle[11pt,emulateapj]{article}
\def\etal{{\sl~et\,al.~}}

\begin{document} 

\title {Young Red Spheroidal Galaxies in the Hubble Deep Fields: 
Evidence for a Truncated IMF at $\sim2M_{\odot}$ and a Constant Space 
Density to z$\sim2$\altaffilmark{1,2,3}}
\altaffiltext{1}{Based on observations made with NASA/ESA Hubble Space
Telescope which is operated by AURA, Inc., under contract with NASA.}
\altaffiltext{2}{ Based on observations made at the Kitt Peak National
Observatory, National Optical Astronomy Observatories, which is
operated by the Association of Universities for Research in Astronomy,
Inc. (AURA) under cooperative agreement with the National Science
Foundation.}
\altaffiltext{3}{Observations have been carried out using the ESO New
Technology Telescope (NTT) at the La Silla observatory under
Program-ID Nos. 59.A-9005(A) and 60.A-9005(A).}

\author
{Tom Broadhurst, Rychard J. Bouwens}
\affil
{Department of Astronomy, University of California, Berkeley, CA 94720}

\begin{abstract}
  
The optical-IR images of the Northern and Southern Hubble Deep Fields
are used to measure the spectral and density evolution of early-type
galaxies. The mean optical SED is found to evolve passively towards a
mid F-star dominated spectrum by z$\sim$2.  We demonstrate with
realistic simulations that hotter ellipticals would be readily visible
if evolution progressed blueward and brightward at z$>$2, following a
standard IMF. The colour distributions are best fitted by a `red' IMF,
deficient above $\sim2M_{\odot}$ and with a spread of formation in
the range 1.5$<z_f<$2.5.  Traditional age dating is spurious in this
context, a distant elliptical can be young but appear red, with an
apparent age $>3$Gyrs independent of its formation redshift. Regarding
density evolution, we demonstrate that the sharp decline in numbers
claimed at z$>$1 results from a selection bias against distant red
galaxies in the optical, where the flux is too weak for morphological
classification, but is remedied with relatively modest IR exposures
revealing a roughly constant space density to z$\sim$2. We point out
that the lack of high mass star-formation inferred here and the
requirement of metals implicates cooling-flows of pre-enriched gas in
the creation of the {\it stellar} content of spheroidal galaxies.
Deep-field X-ray images will be very helpful to examine this
possibility.

\end{abstract}


\section{Introduction}
 
Elliptical galaxies are anomalous in many respects when considered in
the context of the standard ideas regarding galaxy and star-formation.
Despite the absence of star-formation today only minimal passive
evolution has been identified to z$\sim$1, mainly from optical-IR
colours of cluster sequences, which are marginally bluer than
k-corrections predict (Stanford, Eisenhardt \& Dickinson 1998).  At
higher redshift, examples of luminous red galaxies are found with F
and G-star dominated spectra (Dunlop \etal 1996; Spinrad \etal 1998;
Broadhurst \& Frye 1999). No bright blue precursors have been
identified. The absence of precursors naively implies the early epoch
was obscured by dust or restricted to unobservably high
redshifts. Benitez \etal (1999) strongly limit any unobscured
formation to z$>$10, in the deepest available VLT/NICMOS images.
Support for early dust is controversial. Claims of high star-formation
rates from Far-IR imaging of the HDF (Hughes et al 1998) have been
corrected in the radio (Richards 1998) and identified with low
redshift disk galaxies and AGN. This is consistent with the
non-detection of Far-IR emission from targeted observations of
optically luminous $z\sim 3$ galaxies (Scott \etal 1998).

In the context of hierarchical models, it is natural to view E/SO
galaxies as the end product of a merging process and hence to predict
declining numbers with increasing redshift.  Locally at least, merging
of disk galaxies is seen to create some spheroidal shaped objects
(Schweizer \& Seitzer 1992). At faint magnitudes claims have been made
for a decline in the space density of red and/or elliptical galaxies
at z$>$1, (Kauffmann, White, \& Charlot 1996; Zepf 1997; Franceschini
\etal 1997; Kauffmann \& Charlot 1998; Menanteau \etal 1999; Barger et
al.\ 1999) mainly on the basis of optical imaging.  Conflicting with
this is the simplicity of structural and color relations between
elliptical galaxies (Faber \& Jackson 1976), particularly in rich
clusters where coeval monolithic formation is inferred (Bower, Lucey,
\& Ellis 1992).

If gas rich mergers of spirals are to produce the ellipticals then the
enhanced alpha-element abundance generated by a brief merger induced
episode of star-formation is unacceptably diluted by the pre-existing
SNIa-enriched ISM (Thomas, Greggio \& Bender 1999). Furthermore the
mass of stars formed during a merger is limited by the general absence
of an intermediate-age stellar population in post-merger ellipticals
(Silva \& Bothun 1998; James \& Mobasher 1999).  Metallicity is also a
problem with monolithic collapse formation, as closed box
star-formation does not account for the observed lack of low
metallicity stars (Worthey 1994) but implies pre-enrichment of the gas
(Thomas, Greggio \& Bender 1999).

Recently, distant red ellipticals and other spheroid dominated galaxies
at 1$<$z$<$2 have been detected in the deepest combination of
optical-IR images -- a small NICMOS/VLT field (Benitez \etal 1999,
Treu \etal 1998, Stiavelli \etal 1998).  Here we analyse a much larger
sample of distant red galaxies by combining optical-IR photometry from
both Hubble Deep Fields, to measure the rates of spectral and density
evolution (\S3, \S4) with photometric redshift measurements (\S2) and
discuss new implications for their formation (\S4).

\section {Observations}

The observations used here are the deep HST optical images in the UBVI
(Williams et al.\ 1996, 1999) and the JHK images from KPNO (Dickinson
\etal 1997) and from the SOFI instrument on the NTT in the south (da
Costa \etal 1998).  We use the published zero-points, filter
transmission and detector response curves. Magnitudes are measured
using SExtractor (Bertin \& Arnouts 1996), and photometric redshifts
are estimated by maximizing the likelihood with respect to a set of
instantaneous burst spectra calculated using the $[Z/Z_{\odot}] =
-0.2$ Bruzual/Charlot spectrophotometric package (Leitherer et al.\
1996).  Stars are distinguishable to very faint magnitudes by both a
stellarity index (Bertin \& Arnouts 1996) and, interestingly for red
stars, by a poor fit to redshifted red galaxy spectra.

\section{Spectral Evolution}

The first point to note is that the choice of metallicity does not
significantly affect the redshift estimate. The break at 4000\AA\ is
so sharp that with accurate optical-IR magnitudes the redshift can be
determined to $\sim$15\% with a ruler.  A comparison of photo-z's with
the 10 spectroscopic redshifts of elliptical galaxies the HDF is very
good with a hint of a $\delta$z=0.1 systematic overestimate.  48
objects are found with SEDs well matched to early type galaxies.  The
majority clearly fit a de-Vaucouleurs profiles and not an exponential
(Fig 2). The remainder are too faint in the optical and of too low
resolution in the IR to constrain the morphology.  In other words, our
sample contains all objects consistent with a de-Vaucoulours profile
and a passively evolving SED.  Half of these galaxies lie at z$>$1 and
extend to z=2.5.  Few if any moderately bright blue ellipticals ($\sim
5$ at $m_I \sim 23-24$) are missed this way.

The spectra are compared in the restframe (Fig 1). A clear
evolutionary trend emerges towards a mid F-star dominated spectrum by
$z\sim2$.  Hotter A-star dominated spectra would be very easily
recognized if ellipticals evolved further according to a standard IMF
at z$>$2, such young ($<$1Gyr) precursors being very bright and blue.
This simple result suggests that the passive evolution of elliptical
galaxies begins at $\sim2M_{\odot}$. For approximately $\sim$1-2Gyrs
after formation the spectrum of such a stellar population has no
detectable spectral evolution, consistent with the slow evolution
found here above $z\sim1$ (Fig 1). This level of evolution corresponds
to a change in magnitude in the rest-frame B of only $1.^{m}2$ between
z$\sim$2 and the present.  A small but detectable variance among SEDs
is observed at any redshift (Fig 1) with evidence for greater
homogeneity at high redshift.

The bluest 3 objects marking the starting point of the color-colour
tracks in Fig 3 are estimated to lie in the range 1$<$z$<$2. These
objects contain a small blue excess in U and B relative to an F-star
spectrum (Fig 2, bottom panel) which is spatially distributed like the
general light profile (Fig 2), ruling out an AGN
contribution. Accommodating this with some A-star light steepens the
IR appreciably, requiring a redder IMF for a good fit.  Nebular
continuum emission is an interesting possibility. Spectroscopy would
be very helpful in understanding these relatively blue ellipticals.

\section{Density Evolution}

A proper assessment of density evolution requires simulated images to
account for the very strong redshift dependent
k-correction.  Simulations are made in all bands using the local
luminosity function of early type galaxies (Pozzetti et al 1998) and
matched in background noise, pixel scale and PSF of each passband,
using a variation of the machinery described in Bouwens, Broadhurst \&
Silk (1998). Selection and photometry of both the observed and
simulated images is performed identically. Fig 3 shows a comparison
with a model in which the density is fixed and only the observed
minimal spectral evolution takes place to $z_f$=2.5. It is clear that
the numbers and luminosities of red galaxies has not changed much
between z$\sim$2 and the present, in agreement with the claim of
Benitez \etal (1999), but inconsistent with other estimates, in
particular previous optical work.  The unknown volume at high redshift
translates into an uncertainty in the predicted numbers at z$>$1, so
that both low $\Omega$ and flat $\Lambda$ dominated models
underpredict the data by $\sim$30\%.

\section{Discussion and Conclusions}

Our findings show that the passive evolution of ellipticals evolves
slowly to a mid-F star spectrum by $z\sim2$.  Bluer ellipticals are
conspicuous by their absence, and at face value, this simply suggests
that the main sequence in elliptical galaxies does not extend above
$\sim 2 M_{\odot}$.  It is also clear that most elliptical galaxies
form at z$>$1 given the lack of any significant decline in their space
density with redshift, subject to a factor of ~30\% uncertainty from
the unknown volume.  These conclusions are surprising given the high
metal content of ellipticals and implies some gas pre-enrichment.
This requirement is more palatable in light of recent evidence of
outflows in higher redshift star forming galaxies, in particular
lensed galaxies for which there is sufficient signal to detect
blueshifted ISM absorption lines (Franx \etal 1997; Frye \&
Broadhurst; Frye \etal 1999).  Such outflows will be preferentially
enriched with alpha-elements from SNII activity.

Independent of the observed outflows, Renzini (1997) has argued
convincingly that SNII enrichment of the ICM is indicated by the
predominantly alpha-element enriched gas.  In the context of
hierarchical evolution early enriched material will cool onto the
later forming more massive halos. Locally, examples of near solar
enriched cooling-flow X-ray gas is found in groups and clusters of
galaxies centered on giant ellipticals of similar metallicity
(Finoguenov \& Ponman 1999).  We suggest that cooling may be
responsible for the formation of the stellar content of elliptical
galaxies more generally, naturally leading to a bottom heavy IMF
consistent with our results.  Inviting this simple picture is the
remarkable correspondence between the most luminous X-ray cooling gas
with impressively large cD galaxies (Fabian, Nulsen, \& Canizares
1991).  Hence it is perhaps not surprising to find that such objects
contain {\it young} stellar populations (Mehlert \etal 1997) if this
cooling gas forms visible stars.  A clear test of the possible role of
cooling flows in the formation of spheroidal galaxies will be provided
soon by deep field X-ray imaging, like the planned deep AXAF field
(Giacconi \etal 1999). Constraining the numbers of even higher
redshift red galaxies requires deeper IR imaging to explore beyond
z=2. The ISAAC camera on the VLT has the area and efficiency to
achieve this, extending $\sim2$ magnitudes fainter than the relatively
low resolution $\sim$4m IR imaging used here.

\acknowledgments

We thank Piero Rosati and Alvio Renzini for useful conversations. TJB
acknowledges NASA grant AR07522.01-96A.

\fontsize{10}{14pt}\selectfont

\begin{figure}[t] 
\epsscale{1}
\plotone{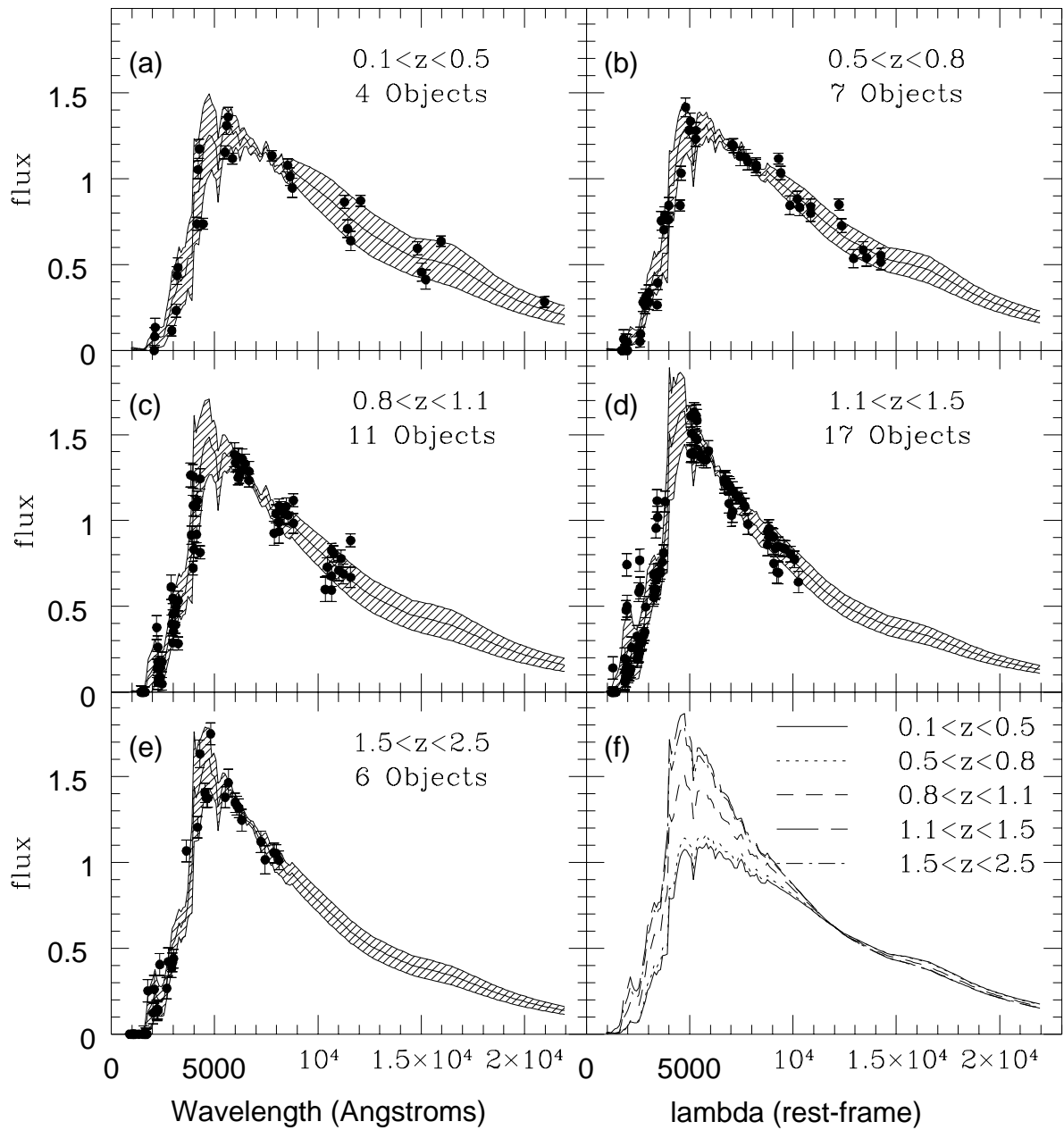}
\caption{The spectral evolution of the early type galaxies can be
traced with redshift via their mean SEDs (a--e) in the range z$<$2.5.
A detectable variance between SEDs is apparent at all redshifts but a
general smooth trend is apparent towards the peaky spectrum of an
F-star dominated spectrum.  Panel (f) shows the trend of the mean
model SED over the full redshift range normalized in the red.  Nothing
bluer than a mid type F-star spectrum is found.}
\end{figure}

\begin{figure}[t] 
\epsscale{0.9}
\plotone{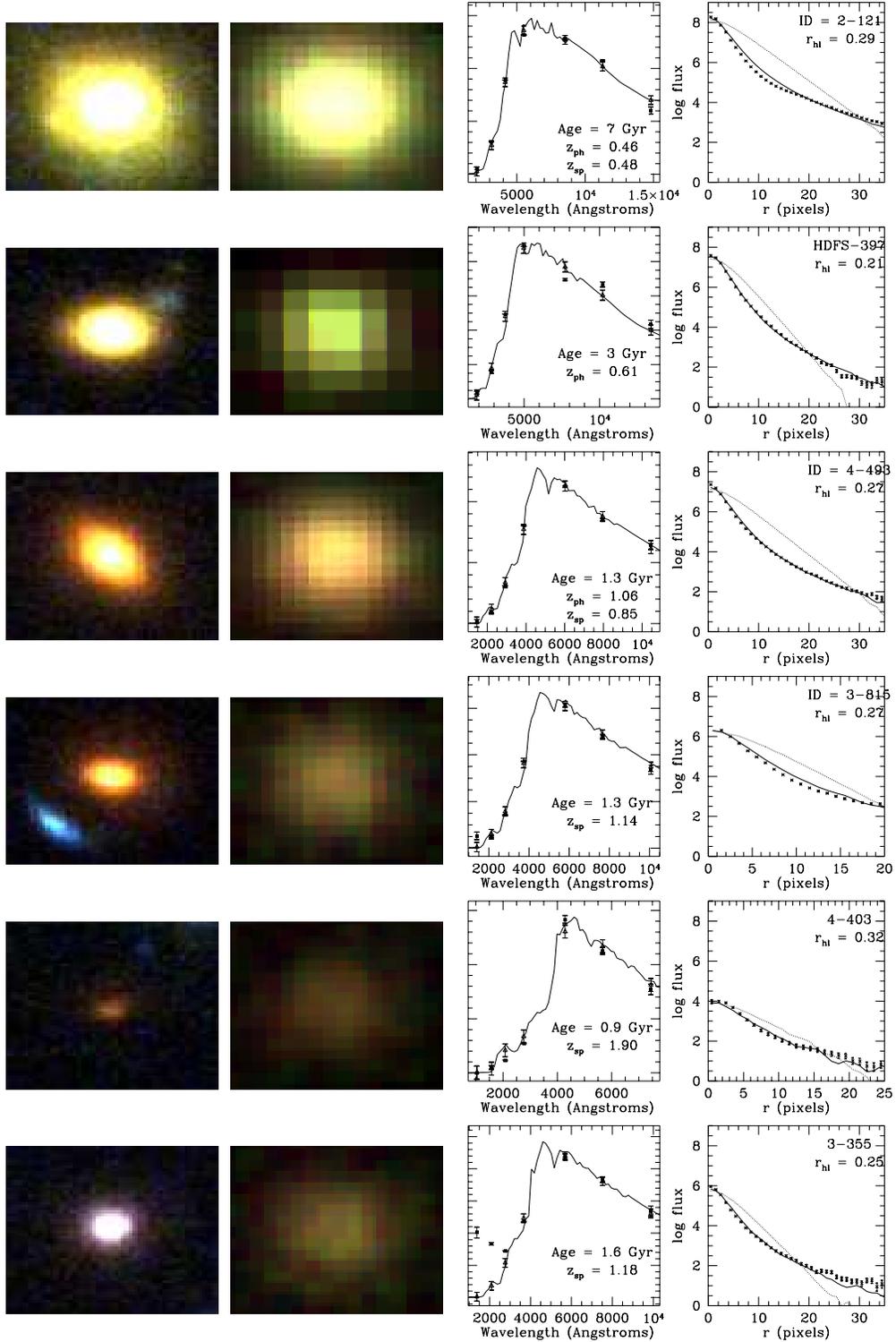}
\caption{Typical spheroidal galaxies as a function of redshift.  We
include the optical colour image, the IR (KPNO or NTT) image, the SED
fits (model fluxes are given by open triangles) to the observations
(square points), and the observed 1-D profiles (crosses with errors)
compared to the best-fit de-Vaucouleur profile (solid line) and
exponential profile (dotted line).  The bottom panel shows an example
of an anomalous morphologically early-type galaxy with a U--B excess
(see text).}
\end{figure}

\begin{figure}[t] 
\epsscale{1.0}
\plotone{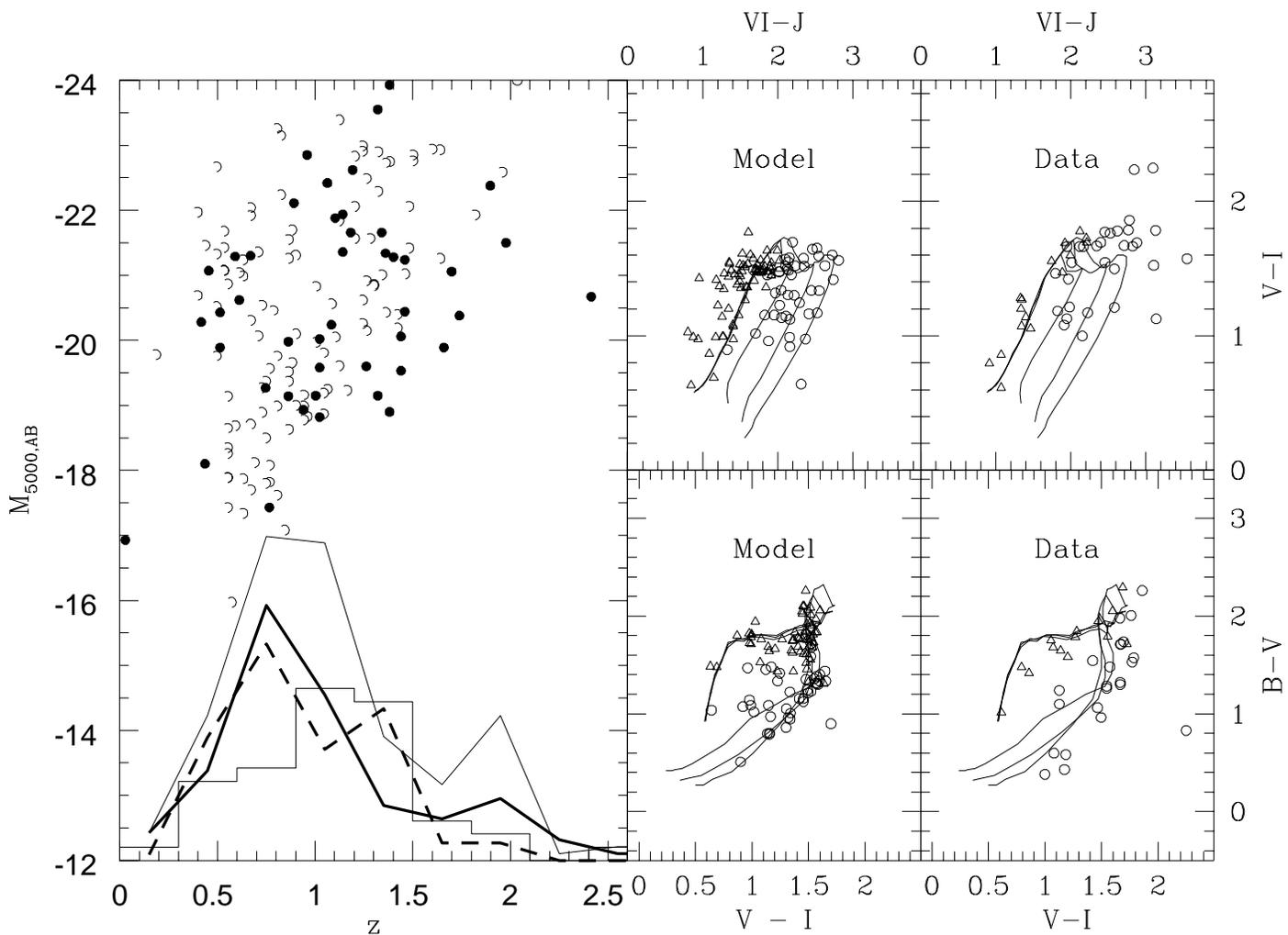}
\caption{ The dot plot on the right compares the redshift and absolute
magnitudes (at 5000 \AA) of the observations (solid circles) with
ellipticals recovered from our simulations (open circles, 2.5x in
density) for our best-fit passive evolution model truncated at
$3M_{\odot}$ (using the Fioc \& Rocca-Volmerange (1997)
spectrophotometric tables) with formation redshifts distributed
between z$=$1.5 and 2.5. The histogram shows the observed redshift
distribution compared with the above best-fit model (dashed line) and
a passive evolution model with $z_f = 3$ for two choices of geometry
($\Omega = 1$, thick line; $\Omega = 0.3$, thin line).  Clearly, at
most a factor of $30\%$ decline in red galaxies is measured for the
large volume models, but no trend to lower luminosity is found with
increasing redshift.  Colour-colour diagrams are also shown with
evolutionary tracks indicating the sensitivity to formation redshift,
$z_f$=1.5,1.75 \& 2, progressing redder for later formation.}
\end{figure}

\end{document}